# Role of α-cluster transfer in formation $^{20}$Ne+$^{16}$O cross sections at backward angles

Sh. Hamada[1], Nourhan M. Elmedalaa[1], I. Bondouk[1], N. Darwish[1]

[1]Faculty of Science, Tanta University, Tanta, Egypt
sh.m.hamada@science.tanta.edu.eg

**Abstract:** The experimental angular distributions for $^{20}$Ne+$^{16}$O elastic transfer are reanalyzed using different forms of potential both phenomenological and semi-microscopic. The significant increase in cross sections at backward hemisphere due to the contribution of α-cluster transfer is investigated using the distorted wave Born approximation (DWBA) method. The spectroscopic amplitude (*SA*) for the configuration $^{20}$Ne→$^{16}$O+α at the different concerned energies is extracted. The agreement between the experimental data and theoretical calculations using the two considered approaches is reasonably good.

## I. Introduction

The effect of a nucleon or group of nucleons transfer between two interacting nuclei and its role in formation the differential cross sections at backward angles is well understood. For decades, many experimental studies of transfer reactions were performed using stable beams and recently using radioactive beams [1-4]. Such transfer reactions attracted a plenary attention, as they could be a useful tool to extract important information about nuclear structure and mechanism of interaction. Alpha cluster transfer is considered one of these tools for probing such information mainly due to its high binding energy as well as its significant role in the rise of cross sections at large angles [5-7]. Optical model (OM) which is one of the most fundamental and acceptable models used over the past years couldn't reproduce the experimental data in the whole angular range when such transfer phenomenon is well exhibited, and consequently different methods such as Distorted Wave Born Approximation (DWBA), Coupled Channel Born Approximation (CCBA), Coupled Reaction Channels (CRC) and Continuum Discretized Coupled Channel (CDCC) could be used to fit the data [8]. In all the aforementioned methods, potential for the different concerned channels we are dealing with is a key information. Firstly, in elastic transfer process where the entrance and exit channel are physically indistinguishable, it is axiomatic to take the same potential for both channels. But if we study a transfer reaction "the entrance and exit channels are completely different", potentials for the two channels should be different. Secondly, the potential itself may be phenomenological optical potential (OP) or a more microscopic potentials such as double folding potential (DFP) and cluster folding (CF) potential. $^{16}$O($^{20}$Ne,$^{16}$O)$^{20}$Ne is an example for elastic transfer which was extensively studied before [9-16] and the contribution of α-cluster transfer is observed in formation the cross sections at backward angles, such effect is well known as "Anomalous large angle backscattering" (ALAS).

In Ref. [9], the angular distributions for $^{20}$Ne+$^{16}$O at $E_{lab}$($^{20}$Ne)=50 MeV were measured for ground and (2$^+$, 4$^+$) $^{20}$Ne excited states, data were analyzed using both OM and CRC. In Ref. [10], excitation functions and angular distributions for $^{16}$O($^{20}$Ne,$^{16}$O)$^{20}$Ne were measured in energy range $E_{c.m.}$ = 14.2 – 24.7 MeV, and at energies ($E_{c.m.}$=24.2 and 24.7 MeV) which are slightly above the Coulomb barrier energy for $^{20}$Ne+$^{16}$O nuclear system, the effect of increasing cross sections at large angles was studied in terms of α-exchange. Y. Kondo et al. [11-13] studied the gross structures observed in $^{20}$Ne+$^{16}$O elastic scattering and the $^{20}$Ne($^{16}$O,$^{12}$C)$^{24}$Mg$_{(g.s.)}$ reaction in the range $E_{c.m.}$ = 22.8 – 38.6 MeV using potentials including a parity dependent real part and angular momentum dependent absorptive part, they reported also that the measured angular distributions can be well described by making the optical potential more surface transparent. In Ref. [14], the gross structure observed in the $^{20}$Ne+$^{16}$O angular distributions was analyzed using a deep optical potential. N. Burtebayev et al. [15] tried to reproduce the experimental data for $^{20}$Ne+$^{16}$O at $E_{lab}$=50 MeV, the data was analyzed using deep real potential with a sum of Woods-Saxon typed surface and volume imaginary potentials. All these previous studies constructed the interaction potential on a phenomenological basis.

Years ago, Y. Yang et al. [16] studied $^{20}$Ne+$^{16}$O in the energy range $E_{c.m.}$ =24.5 – 35.5 MeV. The real part of $^{20}$Ne+$^{16}$O potential was expressed in a folding form derived on the basis of 4α model of $^{16}$O nucleus and α+$^{16}$O model of $^{20}$Ne nucleus, the imaginary potentials were taken as imaginary volume part as well as another absorptive part. The experimental data at these energies showed a significant rise in differential cross sections at backward angles due to the well-known role of "α-cluster transfer". The surprise here was that the authors of this work [16] managed from reproducing the experimental data in the whole angular range without the inclusion the effect of "α-cluster transfer". The question now is: as the marked increase in cross sections at backward hemisphere is well agreed to be due to the contribution of α-cluster transfer so, how the potential used in [16] could fairly reproduce the experimental data without inclusion the effect of α-cluster transfer. The only possible answer lies in the shape of the imaginary potential taken in this work.

In the current work, we aimed to reanalyze $^{20}$Ne+$^{16}$O nuclear system using both pure phenomenological as well as semi-microscopic approaches in order to extract reliable information about both the interaction potential and the spectroscopic amplitude (*SA*) for the configuration $^{20}$Ne→$^{16}$O+α. The paper is organized as follow. In Sec. II CF potential for $^{20}$Ne+$^{16}$O nuclear system potential is discussed. Sec. III is devoted to result and discussion. Summary is given in Sec. IV.

## II . Theoretical Formulation
### A. *Phenomenological analysis for $^{20}$Ne+$^{16}$O elastic transfer*

The experimental angular distributions for $^{20}$Ne+$^{16}$O nuclear system in the energy range $E_{c.m.}$=24.5 – 35.5 [10,11] MeV showed a significant increase in differential cross sections at backward angles due to the contribution of α-cluster transfer between $^{20}$Ne and $^{16}$O "see Fig. 1".

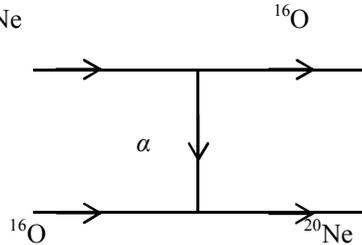

**Fig.1:** $^{16}$O($^{20}$Ne,$^{16}$O)$^{20}$Ne α-cluster transfer process

Theoretical analysis of experimental data was performed firstly on a phenomenological basis within the framework of OM, and the potential parameters were selected to satisfy best fitting between the experimental data and theoretical calculations. Fitting the experimental data was carried out using data in the forward hemisphere "pure elastic scattering" in order to exclude the influence of potential parameters by other mechanisms involved in formation the elastic scattering cross sections. Real and imaginary parts of the potential are taken in Woods-Saxon form factor shape as expressed in Eq. (1).

$$U(R) = V_C - V_0 \left[1 + \exp\left(\frac{r - R_V}{a_V}\right)\right]^{-1} - iW_0 \left[1 + \exp\left(\frac{r - R_W}{a_W}\right)\right]^{-1} \quad (1)$$

Coulomb potential of a uniform charged sphere with radius
$$R_i = r_i(A_P^{1/3} + A_T^{1/3}), \quad i = V, W, C$$

## B. Semi-microscopic analysis for $^{20}$Ne+$^{16}$O elastic transfer reaction

Motivating by the well-known $\alpha+^{16}$O cluster structure for $^{20}$Ne, we tried to reproduce the available experimental data for $^{20}$Ne+$^{16}$O elastic transfer using cluster folding optical model (CFOM). The experimental data taken into consideration are those in energy range where the effect of α-cluster transfer between the interacting nuclei and its role in formation the cross sections at backward angles is well observed. Such calculations were performed using code FRESCO [17]. Within the framework of CFOM, the real part of potential was constructed on the basis of cluster folding in addition to an imaginary potential of the conventional Woods-Saxon (WS) form. The real cluster folding potential used in the current work is presented in Fig. 2.

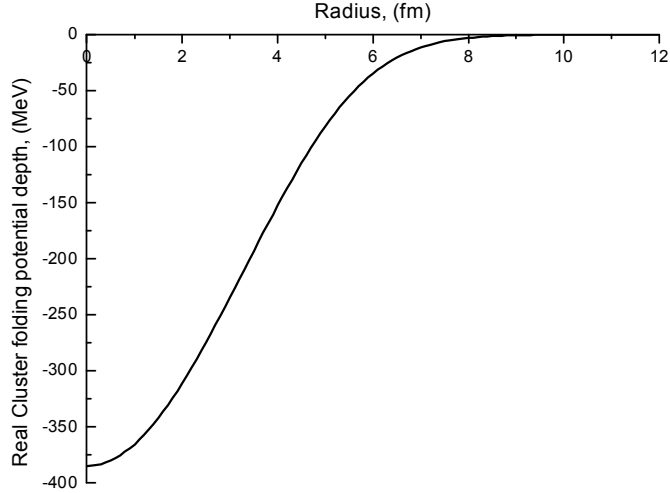

**Fig. 2**: The real cluster folding potential used in the current work is presented

The basic ingredients required to perform CF calculations for $^{20}$Ne+$^{16}$O nuclear system are the potentials for $\alpha+^{16}$O and $^{16}$O+$^{16}$O channels, for simplification we will just call them $V_1$ for ($\alpha+^{16}$O) and $V_2$ for ($^{16}$O+$^{16}$O), as well as binding potential for $\alpha+^{16}$O in $^{20}$Ne nuclei. Each of the two potentials $V_1$ and $V_2$ consists of two parts: real part derived from double folding potential in addition to an imaginary volume part of WS shape. $V_1$ and $V_2$ potential are well tested and they could fairly reproduce the $\alpha+^{16}$O and $^{16}$O+$^{16}$O pure elastic experimental angular distributions at energies $E_{^{16}O} = \left(\frac{16}{20}\right) \times E_{c.m.}$ and $E_\alpha = \left(\frac{4}{20}\right) \times E_{c.m.}$ respectively.

The prepared $^{20}$Ne+$^{16}$O CF potential from folding $V_1$ and $V_2$ was used for fitting the experimental angular distributions data for $^{20}$Ne+$^{16}$O at the forward hemisphere (angles up to 90°) using only two parameters W and $a_W$ (imaginary potential depth and diffuseness), while both $r_W$ (radius parameter for the imaginary volume potential) and $N_{RCF}$ (normalization factor of real cluster folding potential) were kept constant.

The effect of elastic transfer was taken into account in first order by adding the DWBA amplitude $f_{DWBA}(\pi-\theta)$ to the elastic scattering amplitude $f_{el}(\theta)$. The previously obtained optimal parameters from both OM and CFOM calculations were taken constant in DWBA calculations and with taking SA as free parameter. The optimal value for SA is: that produces the least $\chi^2$ value when fitting data at backward angles. In other words, spectroscopic factors SF "which are the square of SA" was considered as phenomenological parameters and determined from a comparison of theoretical and experimental cross sections at backward angles.

In the framework of the proposed cluster folding model based on $\alpha+^{16}O$ cluster structure for $^{20}Ne$ nucleus, the real part of CF potential has the form:

$$V_{CF}(\mathbf{R}) = \int \left[ V_{\alpha\,^{16}O}\left(\mathbf{R} - \frac{4}{5}\mathbf{r}\right) + V_{^{16}O\,^{16}O}\left(\mathbf{R} + \frac{1}{5}\mathbf{r}\right) \right] |\chi_0(\mathbf{r})|^2 d\mathbf{r}, \quad (2)$$

where $\chi_0(\mathbf{r})$ is the wave function for the relative motion of $\alpha$ and $^{16}O$ in the ground state of $^{20}Ne$ [18], and $\mathbf{r}$ is the relative coordinate between the centers of mass of $\alpha$ and $^{16}O$. $V_{\alpha\,^{16}O}$ and $V_{^{16}O\,^{16}O}$ are the effective interaction potentials which are taken to be of the BDM3Y1 form based on the M3Y-Paris potential and consists of two parts direct part $v_D$ and exchange part $v_{EX}$:

$$v_D(s) = 11061.625 \frac{\exp(-4s)}{4s} - 2537.5 \frac{\exp(-2.5s)}{2.5s},$$
$$v_{EX}(s) = -1524.25 \frac{\exp(-4s)}{4s} - 518.75 \frac{\exp(-2.5s)}{2.5s} - 7.8474 \frac{\exp(-0.7072s)}{0.7072s}, \quad (3)$$

The M3Y-Paris interaction is scaled by an explicit density-dependent function $F(\rho)$:

$$v_{D(EX)}(\rho, s) = F(\rho) v_{D(EX)}(s), \quad (4)$$

The density-dependent function can be written as:
$$F(\rho) = C[1 + \alpha \exp(-\beta\rho) - \gamma\rho^n], \quad (5)$$

the parameters $C$, $\alpha$, $\beta$, $\gamma$ for BDM3Y1 [19] are listed in table I. The density distributions of $\alpha$ and $^{16}O$ are expressed in a modified form of the Gaussian shape as $\rho(r) = \rho_0(1 + wr^2)\exp(-\beta r^2)$, where $\rho_0$=0.1317, $w$=0.6457 and $\beta$=0.3228 for $^{16}O$ [20] and $\rho_0$=0.4229, $w$=0.0 and $\beta$=0.7024 for $\alpha$-particles [21].

**Table I:** Parameters of density-dependent function $F(\rho)$

| Interaction Model | c | α | β (fm$^3$) | γ (fm$^{3n}$) | n | K (MeV) |
|---|---|---|---|---|---|---|
| **BDM3Y1** | 1.2521 | 0.0 | 0.0 | 1.7452 | 1 | 270 |

## III. Results and discussion
### A. *Elastic and elastic transfer $^{20}Ne+^{16}O$ analysis using OM*

The comparison between the experimental data and the phenomenological theoretical calculations for $^{20}Ne+^{16}O$ nuclear system at the different concerned energies is shown in Fig. 3.

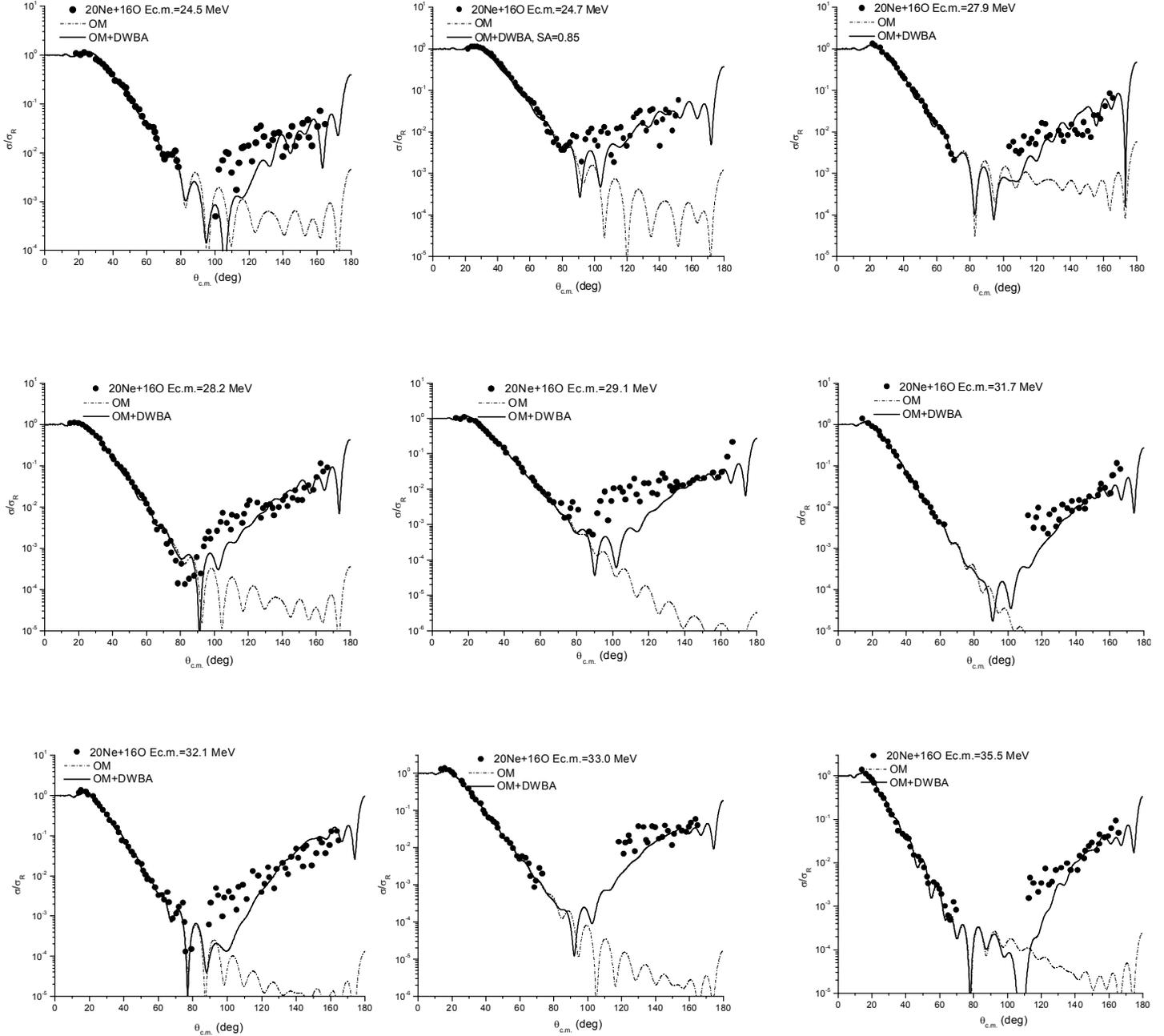

**Fig. 3**: Comparison between the experimental data and the theoretical calculations for $^{20}Ne+^{16}O$ nuclear system at energies $E_{c.m.}$ (24.5, 24.7, 27.9, 28.2, 29.1, 31.7, 32.1, 33.0 and 35.5 MeV) respectively using both (OM) and (DWBA)

At backward angles, there is a significant increase in cross sections due to the contribution of *α*-cluster transfer. Solid black lines represent DWBA calculations after inclusion the role of *α*- cluster transfer. The optimal optical potential parameters obtained from the analysis of pure elastic scattering were used in analysis elastic transfer in addition to the core potential, and the spectroscopic factor was taken as a free parameter. The bound state for the cluster with core used Woods-Saxon potential with radius $R=1.4 \times A^{1/3}$ ($A=16$), the diffusivity of the well was fixed at $a=0.65$fm and the potential depth was adjusted to reproduce the binding energy for the cluster (for B.E=4.729 MeV) at $V_0=114$ MeV. Cluster state quantum numbers were determined using Talmi–Moshinsky formula $2(N-1)+L = \sum_{i=1}^{n} 2(n_i-1)+l_i$, where $n_i, l_i$ are quantum numbers of components of a cluster of nucleons in harmonic oscillator model, $N, L$ are the cluster quantum numbers. Cluster quantum numbers for the overlap used in our calculations are listed in table II.

**Table II** – $N, L, S$ and $J$ for the overlap used in our calculations

| Overlap | N<br>No. of nodes | L | S | J=L+S | B.E<br>MeV |
|---|---|---|---|---|---|
| $\langle ^{20}\text{Ne} \mid ^{16}\text{O} \rangle$ | 5 | 0 | 0.0 | 0.0 | 4.729 |

Optimal OM parameters for $^{20}$Ne+$^{16}$O nuclear system as well as the extracted spectroscopic amplitude are listed in tables **III**. Radii parameters for Coulomb, real and imaginary parts of potential - $r_C$, $r_V$, $r_W$ - were fixed at 1.3, 1.0, 1.25 fm respectively, and diffuseness for the imaginary part of the potential ($a_W$) was also fixed at 0.7 fm. In other words, the data are fitted within the framework of OM utilizing only three parameters: real ($V$) and imaginary ($W$) potential depths as well as the diffuseness ($a_W$) for the imaginary part of the potential.

**Table III** – Optimal optical potential parameters for $^{20}$Ne+$^{16}$O nuclear system at different energies, values of the SA as well as real and imaginary volume integral are also listed. $r_V$ $r_W$ and $a_W$ are fixed at 1.0, 1.25 fm and 0.7 fm respectively

| $E_{c.m.}$<br>MeV | Model | V<br>MeV | $a_V$<br>fm | W<br>MeV | $\chi^2/N$ | SA | $J_V$<br>MeV.fm$^3$ | $J_W$<br>MeV.fm$^3$ | $\sigma$<br>(mb) |
|---|---|---|---|---|---|---|---|---|---|
| 24.5 | OM | 118.9 | 0.698 | 7.07 | 1.2 | | 262.3 | 28.87 | 1157 |
| | OM+DWBA | 118.9 | 0.698 | 7.07 | 35.1 | 0.816 | | | |
| 24.7 | OM | 124.1 | 0.671 | 8.27 | 6.3 | | 270.7 | 33.75 | 1171 |
| | OM+DWBA | 124.1 | 0.671 | 8.27 | 57.7 | 0.85 | | | |
| 27.9 | OM | 113.8 | 0.683 | 7.23 | 4.3 | | 249.5 | 29.5 | 1258 |
| | OM+DWBA | 113.8 | 0.683 | 7.23 | 36.3 | 0.85 | | | |
| 28.2 | OM | 98.1 | 0.729 | 8.9 | 1.6 | | 219.4 | 36.32 | 1315 |
| | OM+DWBA | 98.1 | 0.729 | 8.9 | 24.8 | 0.85 | | | |
| 29.1 | OM | 98.0 | 0.734 | 13.0 | 1.1 | | 219.7 | 53.05 | 1418 |
| | OM+DWBA | 98.0 | 0.734 | 13.0 | 14.2 | 0.85 | | | |
| 31.7 | OM | 87.84 | 0.77 | 15.2 | 4.9 | | 200.1 | 62.02 | 1507 |
| | OM+DWBA | 87.84 | 0.77 | 15.2 | 22.9 | 0.85 | | | |
| 32.1 | OM | 80.3 | 0.73 | 8.96 | 1.8 | | 179.7 | 36.56 | 1388 |
| | OM+DWBA | 80.3 | 0.73 | 8.96 | 18.9 | 1.1 | | | |
| 33 | OM | 86.5 | 0.677 | 10.6 | 0.77 | | 189.2 | 43.25 | 1423 |
| | OM+DWBA | 86.5 | 0.677 | 10.6 | 24.3 | 0.85 | | | |
| 35.5 | OM | 91.95 | 0.724 | 9.54 | 5.9 | | 205.2 | 38.93 | 1479. |
| | OM+DWBA | 91.95 | 0.724 | 9.54 | 24.3 | 0.85 | | | |

## A. Elastic and elastic transfer $^{20}$Ne+$^{16}$O analysis using CFOM

The comparison between experimental data at the different concerned energies and theoretical calculations using CFOM is shown in Fig. 4. The real part of interaction potential created on the basis of cluster folding is unrenormalized "$N_{RCF}$" renormalization factor for the real cluster folding part is fixed to unity". The imaginary part of potential which simply takes into account the reduction of the flux due to absorption was taken to be of Woods-Saxon form factor shape, the total potential used to describe $^{20}$Ne+$^{16}$O is:

$$U(R) = V_C - N_{RCF} V^{CF}(R) - iW(R) \qquad (6)$$

The experimental data were fitted using only two parameters – $W$, $a_W$ –imaginary potential depth and diffuseness respectively, while the real cluster folding part was fixed "$N_{RCF}$=1" and the radius parameter for the imaginary WS potential was fixed at 1.25 fm.

**Table IV** – CFOM parameters for $^{20}$Ne+$^{16}$O nuclear system at different energies, values of the extracted spectroscopic amplitudes as well as real and imaginary volume integral are also listed. $N_{RCF}$ and $r_W$ are fixed at 1"unnormalized" and 1.25 fm respectively.

| $E_{c.m.}$ MeV | Model | $W$ MeV | $a_W$ fm | $\chi^2/N$ | SA | $J_V$ MeV.fm$^3$ | $J_W$ MeV.fm$^3$ | $\sigma$ (mb) |
|---|---|---|---|---|---|---|---|---|
| 24.5 | CFOM | 13.96 | 0.86 | 4.8 | | 437.6 | 59.9 | 1513 |
| | CFOM+DWBA | 13.96 | 0.86 | 27.7 | 1.2 | | | |
| 24.7 | CFOM | 14.29 | 0.694 | 5.1 | | 437.6 | 58.2 | 1305 |
| | CFOM+DWBA | 14.29 | 0.694 | 52.3 | 1.2 | | | |
| 27.9 | CFOM | 13.39 | 0.801 | 5.4 | | 437.6 | 56.4 | 1548 |
| | CFOM+DWBA | 13.39 | 0.801 | 36.0 | 0.9 | | | |
| 28.2 | CFOM | 15.68 | 0.922 | 3.5 | | 437.6 | 68.8 | 1800 |
| | CFOM+DWBA | 15.68 | 0.922 | 20.9 | 1.21 | | | |
| 29.1 | CFOM | 17.1 | 0.786 | 0.43 | | 437.6 | 71.6 | 1628. |
| | CFOM+DWBA | 17.1 | 0.786 | 13.5 | 1.13 | | | |
| 31.7 | CFOM | 17.79 | 0.912 | 5.4 | | 437.6 | 77.7 | 1933 |
| | CFOM+DWBA | 17.79 | 0.912 | 16.8 | 1.29 | | | |
| 32.1 | CFOM | 13.51 | 0.805 | 1.8 | | 437.6 | 56.9 | 1388 |
| | CFOM+DWBA | 13.51 | 0.805 | 30.0 | 1.2 | | | |
| 33 | CFOM | 24.95 | 0.677 | 6.2 | | 437.6 | 101.1 | 1654 |
| | CFOM+DWBA | 24.95 | 0.677 | 38.2 | 1.12 | | | |
| 35.5 | CFOM | 17.68 | 0.816 | 6.2 | | 437.6 | 74.8 | 1841 |
| | CFOM+DWBA | 17.68 | 0.816 | 19.4 | 1.26 | | | |

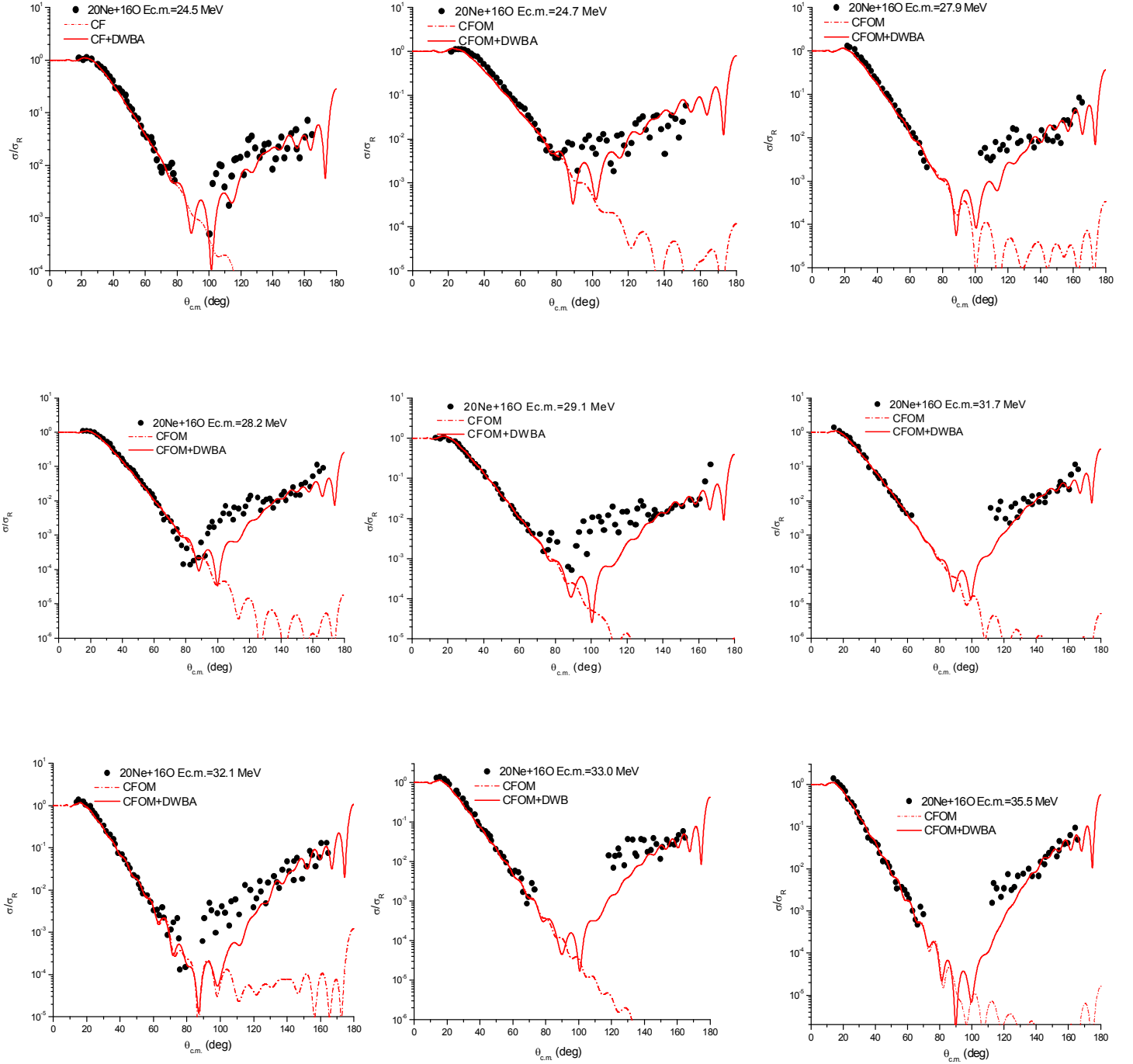

**Fig. 4**: Comparison between the experimental data and the theoretical calculations using CFOM and DWBA for $^{20}Ne+^{16}O$ nuclear system at energies $E_{c.m.}$ (24.5, 24.7, 27.9, 28.2, 29.1, 31.7, 32.1, 33.0 and 35.5 MeV) respectively using total potential of Eq. (6).

The average extracted value for *SA* from OM+DWBA and CFOM+DWBA calculations are ≈0.87±0.09 and ≈1.17±0.11 respectively which are in agreement with the recently reported value 0.72 [22]. In comparison with previously reported values [23-26], it seems that, the value of *SA* is highly potential dependent. The imaginary volume integral ($J_W$) ranges from 28.87 to 62.02 MeV.fm$^3$ and increases with increasing energy and real volume integral ($J_V$) ranges from 179.7 to 270.7 MeV.fm$^3$ and decreases with increasing energy (see Table III). As shown in Fig. 4, the energy dependence of real volume integral could be expressed as $J_V$ = 445.95-7.6 $E_{c.m.}$, and for imaginary volume integral $J_W$ = 2.2+1.28 $E_{c.m.}$. Extracted reaction cross sections $\sigma_R$ from both OM and CFOM calculations are presented in Fig. 5. It is clearly shown that, reaction cross sections increase with increasing energy and could be approximated as: $\sigma_R$ = 711.3+30.8 $E_{c.m.}$ (from CFOM calculations) and $\sigma_R$ = 429.8+30.9 $E_{c.m.}$ (from OM calculations).

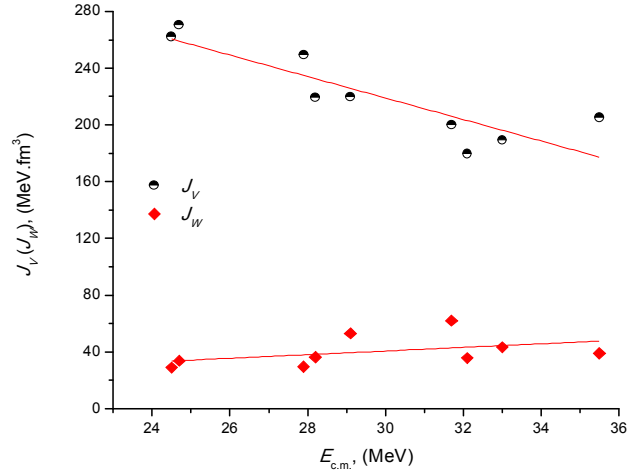

**Fig. 4**: Energy dependence on real and imaginary volume integral extracted from OM calculations

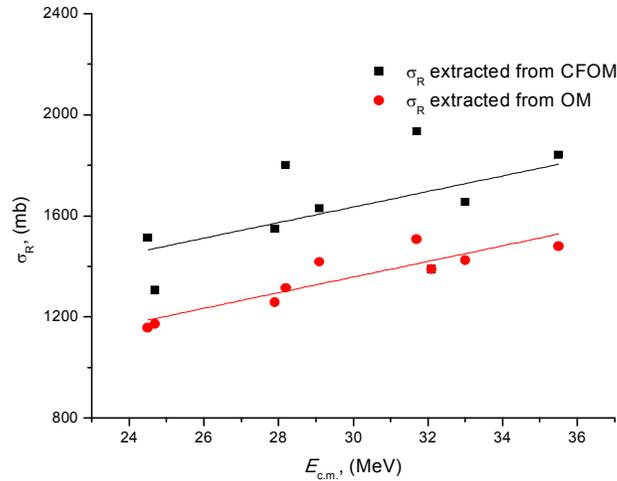

**Fig. 5:** The variation of reaction cross sections with energy extracted from OM and CFOM calculations

## IV. Summary

The available experimental data for $^{20}$Ne+$^{16}$O nuclear system in the energy range $E_{c.m.}$= 24 – 35.5 MeV are reanalyzed using two different potentials: pure OM as well as CFOM potential. In this energy range, the data showed unmistakable rise in cross sections at backward angles. The data was reanalyzed firstly using OM, the optimal potential parameters obtained from analysis of pure elastic scattering were used to analysis the elastic transfer, and the spectroscopic factor was taken as free parameter. Using such method, the agreement between the experimental data and the theoretical calculations were acceptable. Secondly, the concerned data are subjected to an investigation within the framework of CFOM. In this model, the real part of $^{20}$Ne+$^{16}$O interaction potential is constructed on the basis of cluster folding without any renormalization ($N_{RCF}$ = 1) in addition to an imaginary part of potential has a WS shape. The optimal CFOM parameters resulted from fitting the data at forward hemisphere (angles > 90$^o$) are adopted in DWBA to reproduce the data in the whole angular range. The two considered approaches give a reasonable fitting to the experimental data. The average extracted *SA* from OM+DWBA and CFOM+DWBA calculations are ≈0.87±0.09 and ≈1.17±0.11 respectively which are in agreement with the recently reported value 0.72 [22]. These calculations also give an evidence for the appearance of $α$+$^{16}$O structure in the ground state of $^{20}$Ne.